\documentclass[10pt,prc,aps,psfig,twocolumn]{revtex4}
\usepackage{graphicx}
\tighten
\begin{document}
\draft
\title{ Temperature dependence of single-particle properties in nuclear and neutron matter 
in the Dirac-Brueckner-Hartree-Fock model              
 }              
\author{            
 Francesca Sammarruca}      
\affiliation{                 
 Physics Department, University of Idaho, Moscow, ID 83844-0903, U.S.A  } 
\date{\today} 
\email{fsammarr@uidaho.edu}
\begin{abstract}
The understanding of the interaction of nucleons in nuclear and neutron-rich matter at non-zero temperature is  
important for a variety of applications ranging from heavy-ion collisions to nuclear 
astrophysics.                                                                        
In this paper we apply the Dirac-Brueckner-Hartree-Fock method along with the Bonn B potential to predict
single-particle properties in symmetric nuclear matter and pure neutron matter at finite temperature.
It is found that temperature effects are generally small but can be significant at
low density and momentum. 
\end{abstract}
\maketitle

\section{Introduction} 
                                                                     
The nuclear equation of state (EoS) at finite temperature is of fundamental importance for   
heavy-ion (HI) physics \cite{Daniel05,MSU}                                                        
as well as nuclear astrophysics, particularly in the                   
final stages in the evolution of a supernova.                                   
Knowledge of the finite-temperature EoS can be of great help in the interpretation of
experiments aimed at identifying a liquid-gas phase transition.
Presently, findings concerning such transition are very model dependent                                  
\cite{Zuo04,SMN89,JMZ83,Baldo95,JMZ84,HM87,HWW98}. 

When other aspects, such as spin asymmetries of nuclear/neutron matter are included as well, conclusions are
even more contradictory. 
For instance, the influence of finite temperature on the manifestation of 
ferromagnetic instabilities is an unsettled question. In Ref.~\cite{pol23}, phenomenological
Skyrme-type interactions are used in a Hartree-Fock scheme. It is found that the critical
density for ferromagnetism decreases with temperature. On the other hand, 
in Ref.~\cite{temp2} the authors report no indication of ferromagnetic instability at any 
density or temperature based on the Brueckner-Hartree-Fock  (BHF) approximation and the Argonne V18 nucleon-nucleon (NN) interaction \cite{V18}. 
The properties of spin-polarized neutron matter at finite temperature are 
studied in Ref.~\cite{Lopez06} with two different parameterizations of the Gogny interaction. The results show two qualitatively different behaviors for the two parameterizations.
The reasons
for these discrepancies must be carefully studied and their origin understood in terms of specific features
of the nuclear force 
and/or the chosen many-body framework.

 Previous work on the temperature dependence of the EoS includes the calculations by 
Baldo and Ferreira who used the Bloch-De Dominicis diagrammatic expansion \cite{temp1}, Brueckner-Hartree-Fock
calculations with 
and withouth 3BF \cite{Zuo04}, and the predictions of Ref.~\cite{FM03} 
based on the Green's function method.                            
The entropy per particle in symmetric nuclear matter has been studied in 
Ref.~\cite{Rios06} within the 
self-consistent Green's function (SCGF) approach, where                                
both particle-particle and hole-hole scatterings are included. The SCGF framework allows direct access to the 
single-particle spectral function and thus to all the one-body properties of the system. 
A most recent work by Rios {\it et al.} \cite{Rios09} addresses hot neutron matter within the same approch 
and performs a comparison with other models. 
Hot asymmetric matter and $\beta$-stable matter have been studied by Moustakidis {\it et al.}
\cite{Moust08,Moust09} using temperature and momentum dependent effective interactions. 

 The study of the many EoS-related aspects in both symmetric and neutron matter starting from realistic
NN forces and within a microscopic model is still a considerable challenge. 
It is the purpose of this paper to report 
the first part of a comprehensive study of temperature dependence of nuclear and neutron-rich matter 
properties based on the Dirac-Brueckner-Hartree-Fock (DBHF) approach.                                                
An earlier calculation with the DBHF method can be found in Ref.~\cite{HM87}. However, our work will go beyond those
predictions, as we plan to address additional aspects such as (spin/isospin) asymmetric matter and temperature dependence of in-medium effective cross sections. 
Previously, we have confronted isospin asymmetries \cite{AS03,SL08,SL09} as well as spin asymmetries \cite{SK07} effects
on the equation of state of cold matter. 
The inclusion of temperature dependence and  the simultaneous consideration of all of the above mechanisms 
 will make our microscopic predictions more broadly useful and capable to reach out                                         
to the properties of the hot environment present in the latest stage of a supernova collapse  
or in the collision of heavy nuclei at intermediate energies.

Next, after a review of the main aspects of the formalism, 
we will show and discuss predictions of the chemical potential and                                    
single-particle properties in symmetric and pure neutron matter.               
We will concentrate here on the properties of the single-particle within the nuclear medium for the following reason: although the 
nuclear/neutron matter energy density is certainly an important quantity,                                             
the single-particle interaction and its temperature dependence, 
which determine the one-body properties in the medium,      
are perhaps more relevant                                                                 
for non-equilibrium processes such as relativistic
heavy-ion collisions.                                                                  

Hot nuclear and neutron matter are infinite fermionic systems where only the strong interactions among nucleons are 
taken into account. 
The temperatures typically considered are of a few tens of MeV, relatively small on the scale of
nuclear energies. For instance, at densities near the saturation density of nuclear matter, the free Fermi energy, $e_F$, is approximately 40-50 MeV, and thus 
a temperature of 20 MeV is still somewhat low, with $T/e_F \sim 0.5$. 
\section{Formalism }

Within the DBHF method, 
the interactions of the nucleons with the nuclear medium are expressed as self-energy corrections to the 
nucleon propagator. That is, the nucleons are regarded as ``dressed" particles, essentially a gas of non-interacting
quasi-fermions. The behavior of the dressed nucleon is determined by the effective nucleon propagator, which obeys
the Dyson equation.                                                                                                   
Relativistic effects lead to an intrinsically density-dependent interaction which is consistent 
with the contribution from three-body forces (TBF) typically employed in non-relativistic approaches.
The advantage of the DBHF approximation is the absence of phenomenological TBF to be extrapolated
at higher densities from their values determined through observables at normal density. 

In the quasi-particle approximation, the transition to the temperature-dependent case 
is introduced by replacing the zero-temperature
occupation number with its finite-temperature counterpart, namely the Fermi-Dirac occupation density.                       
In fact, it can be shown that the zero-temperature terms in the 
Bethe-Goldstone expansion where temperature is introduced in the occupation number only               
are the dominant ones \cite{temp1}, thus justifying this simplified procedure.
More specifically, one replaces                                                            
\begin{equation}
n(k,\rho) =\left\{
\begin{array}{l l}
1 & \quad \mbox{if $k\leq k_F$}\\
0 & \quad \mbox{otherwise,}
\end{array}
\right.
\end{equation}
with the Fermi-Dirac distribution
\begin{equation}
n_{FD}(k,\rho,T) = \frac{1}{1+e^{(\epsilon(k,\rho,T)-\mu(\rho,T))/T}}. 
\end{equation}
Here T is the temperature in MeV,
$\epsilon(k, \rho, T)$ the single-particle energy, and $\mu$ the chemical potential to
be determined. 
The angle-averaged Pauli operator is evaluated numerically. It's given by
\begin{displaymath}
 Q(q,P,\rho,T) = \frac{1}{2} \int_{-1}^{+1} d(cos~\theta)(1 - n_{FD}(k_1,\rho,T)) \times
\end{displaymath}
\begin{equation}
\times (1 - n_{FD}(k_2,\rho,T))
\end{equation}
for two nucleons having momenta ${\vec k}_1$ and 
${\vec k}_2$, with relative and total momentum 
given by            
${\vec q} =\frac{{\vec k_{1}} -{\vec k_2}}{2}$ and 
${\vec P} = {\vec k}_1 + {\vec k}_{2}$, respectively.

The single-particle energy, $\epsilon(k,\rho,T)$, is now temperature dependent. It can be obtained self-consistently
with the Dirac states 
following the same procedure as used in the zero temperature case, but including Eq.~(2)  
in the calculation of the single-particle potential.  At each iteration of the 
self-consistent calculation, the normalization condition 
\begin{equation}
\rho = D\frac{1}{(2 \pi)^3} \int_0^{\infty} n_{FD}(k,\rho,T)d^3k 
\end{equation}
allows to extract the microscopic chemical potential, $\mu(\rho,T)$. D is a degeneracy factor, equal to 4 
for symmetric unpolarized nuclear matter or 2 for unpolarized neutron matter. 

In close analogy with the $T=0$ case, 
the single-particle potential and the self-consistent nucleon-nucleon G-matrix are related by 
\begin{displaymath}
 U({\vec k},\rho,T) =                                                                  
 \sum_{I,L,S,J} \frac{(2I+1)(2J+1)}{(2t+1)(2s+1)} \times                        
\end{displaymath}
\begin{equation} 
 \times \int _0 ^{\infty} n_{FD}(k',\rho,T)~G^{T,L,S,J}_{N N}(q({\vec k},{\vec k'}),  
 P({\vec k},{\vec k'})) d^3 k', 
\end{equation}
where $I$, $L$, $S$, and $J$ are the NN system quantum numbers while $s$ and $t$ are the single-particle
spin and isospin, respectively. 
Integrations over the Fermi sea are performed using a cutoff adjusted to the density. 
We found a value of 6 fm to be sufficient for the highest densities and temperatures considered here. 

The single-particle energy is the sum of potential ($U$) and kinetic ($KE$) energy contributions, 
\begin{equation}
\epsilon (k,\rho,T) = U(k,\rho,T) + KE(k,\rho,T). 
\end{equation}
It is parametrized using the same {\it ansatz} 
as in the $T=0$ case \cite{Mac89} for a spin-symmetric, rotationally invariant system: 
\begin{equation}
\epsilon(k,\rho,T) = \sqrt{k^2 + (m^*)^2} + U_V, 
\end{equation}
where $U_V$ is the time-like part of the vector potential and $m^*=m+U_S$, with $U_S$ the scalar potential, 
now both density and teperature dependent. This quantity plays an important role in temperature             
and momentum-dependent transport models of heavy-ion collisions.

Once a self-consistent solution is obtained for the single-particle potential, and, simultaneously, for 
the chemical potential, the entropy/particle can be calculated, along with all thermodynamic quantities. 
In the mean-field approximation, the entropy/particle is given by
\begin{displaymath}
S = -\frac{D}{(2 \pi)^3} \int_0^{\infty} [n_{FD}(k,\rho,T)~ln~n_{FD}(k,\rho,T) +        
\end{displaymath}
\begin{equation}
+(1 - n_{FD}(k,\rho,T))~ln~(1 - n_{FD}(k,\rho,T))]~ d^3k, 
\end{equation}
that is, it has the same functional form as in a noninteracting system.

\begin{figure}[!t] 
\centering          
\includegraphics[totalheight=2.7in]{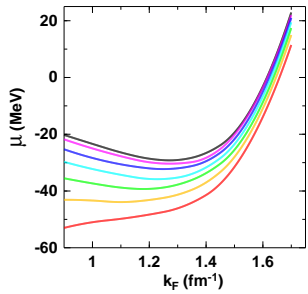}  
\vspace*{-1.2cm}
\caption{(color online)                                        
The chemical potential in symmetric matter as a function of the Fermi momentum at various temperatures 
from $T=0$ to $T=30$ MeV in steps of 5 MeV. The chemical potential decreases with temperature. 
} 
\label{one}
\end{figure}

\begin{figure}[!t] 
\centering          
\includegraphics[totalheight=2.1in]{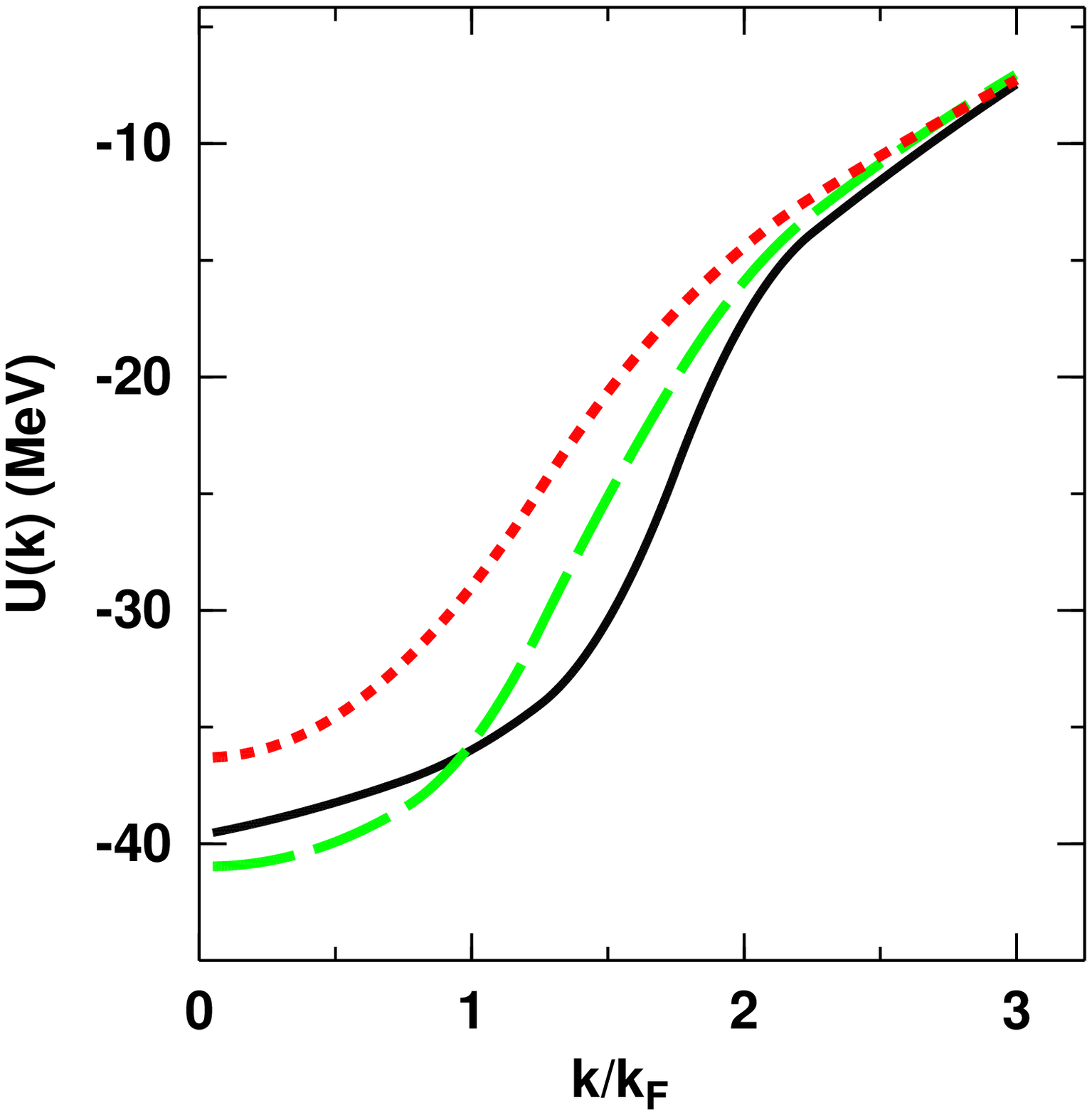}  
\includegraphics[totalheight=2.2in]{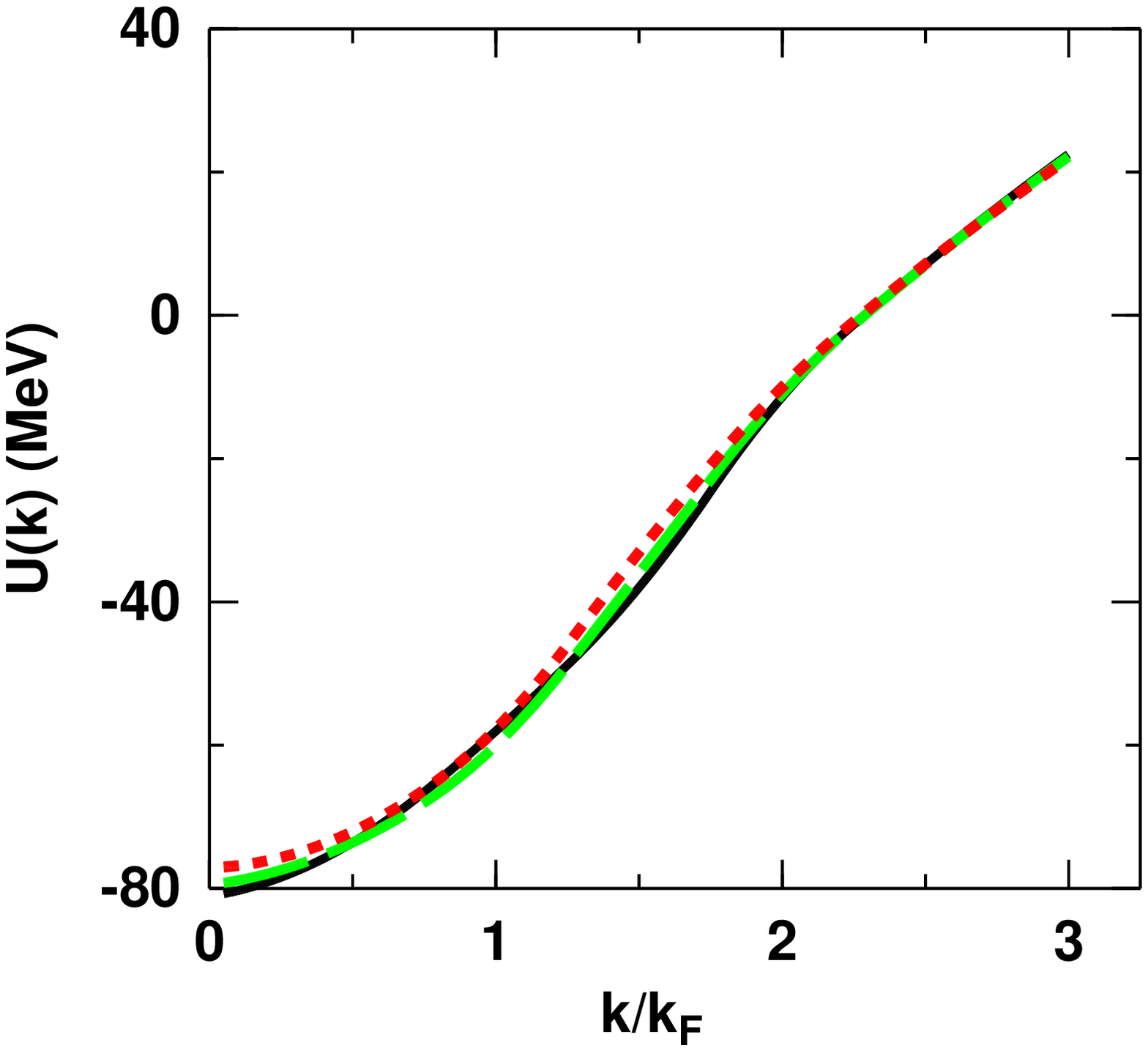}  
\vspace*{-1.2cm}
\caption{(color online)                                        
The single-particle potential in symmetric matter as a function of the momentum for three
different temperatures: $T=0$ (solid black); $T=10$ (dashed green); $T=20$ (dotted red).
The left and right panels correspond to Fermi momenta equal to 0.9 $fm^{-1}$ and
1.3 $fm^{-1}$, respectively. 
} 
\label{two}
\end{figure}

\begin{figure}[!t] 
\centering          
\includegraphics[totalheight=2.7in]{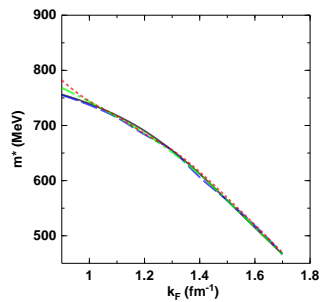}  
\vspace*{-1.5cm}
\caption{(color online)                                        
Effective mass in symmetric matter as a function of the Fermi momentum 
and for different temperatures: $T=0$ (solid black); $T=10$ (dashed blue); $T=15$ (dash-dotted green); 
$T=20$ (dotted red).
} 
\label{three}
\end{figure}

\begin{figure}[!t] 
\centering          
\includegraphics[totalheight=2.1in]{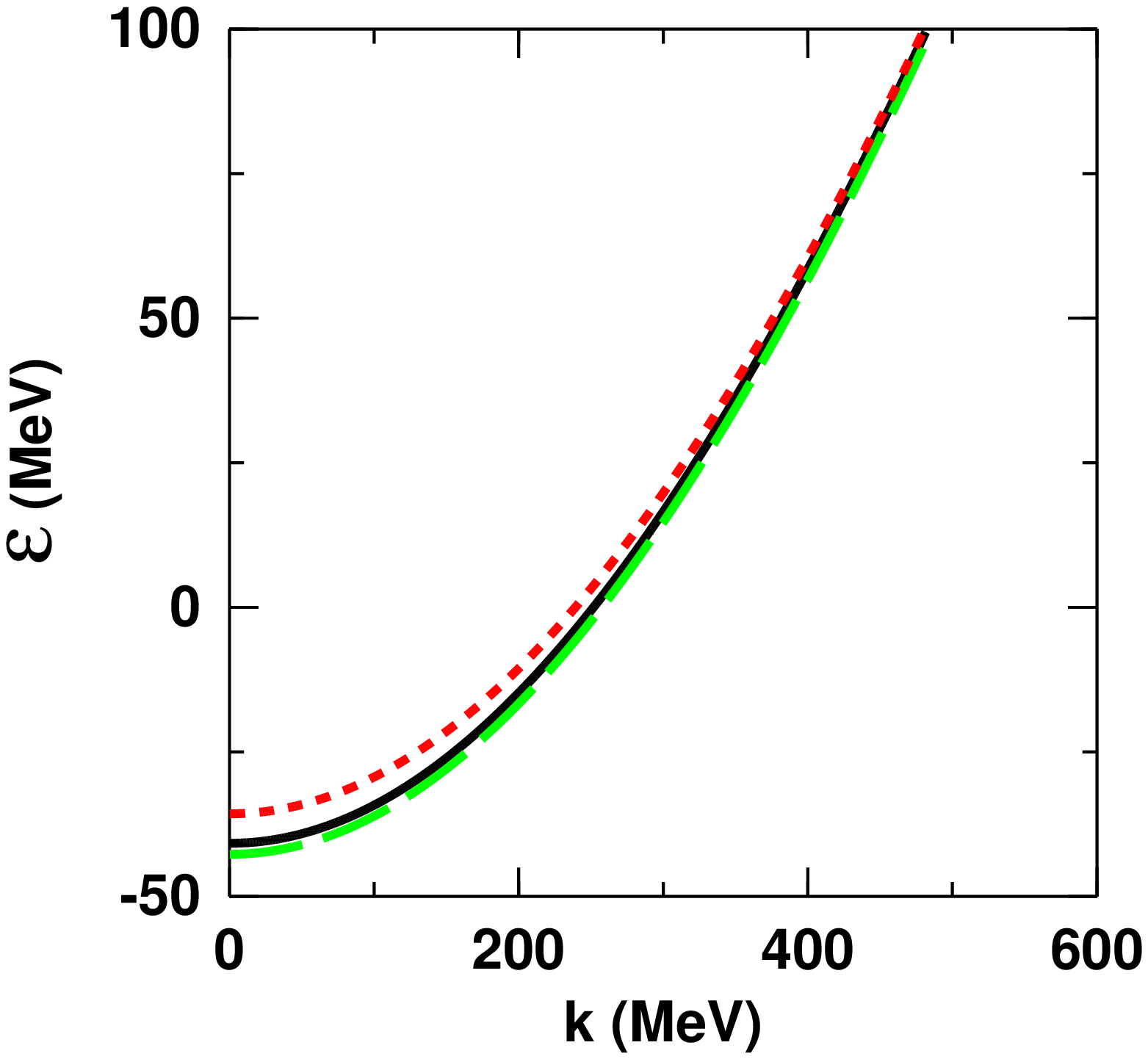}  
\includegraphics[totalheight=2.1in]{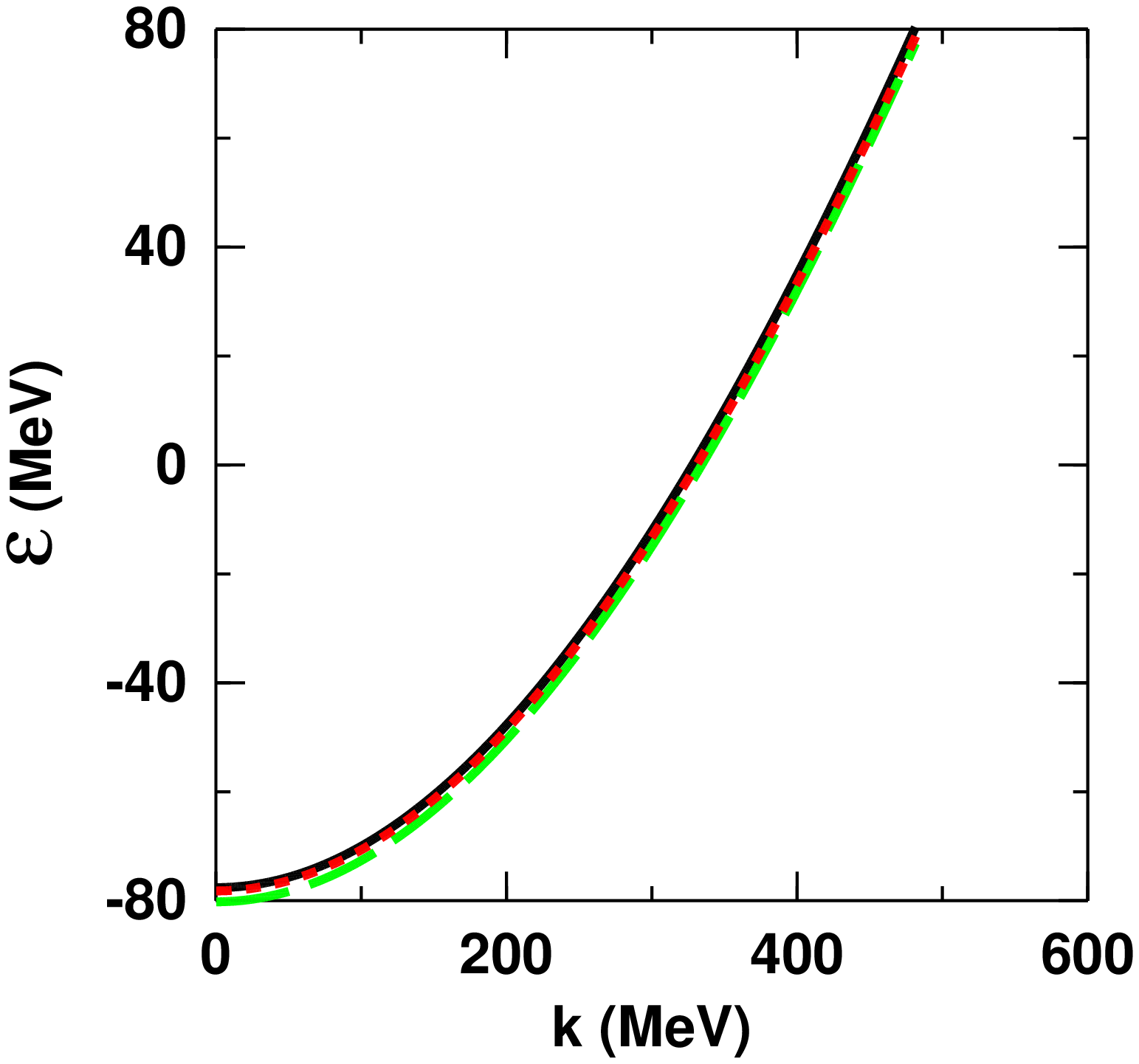}  
\vspace*{-1.2cm}
\caption{(color online)                                        
The single-particle energy minus the rest mass in symmetric matter parametrized as in Eq.~(7) {\it vs.} the momentum at 
different temperatures: $T=0$ (solid black); $T=10$ (dashed green); $T=20$ (dotted red).
The left and right panels correspond to Fermi momenta equal to 0.9 $fm^{-1}$ and
1.3 $fm^{-1}$, respectively. 
} 
\label{four}
\end{figure}
\section{Results}                                                                  
\subsection{Symmetric Nuclear Matter} 
We begin by showing the predicted chemical potential in symmetric nuclear matter (SNM), see Fig.~1. 
Each curve is an isotherm starting from zero temperature and going up to $T=30$ MeV in steps of 5 MeV ($\mu$ goes down with 
increasing temperature). The effect of temperature is definitely larger at the lower densities and
increases in size with increasing temperature. 
Our predictions are in fair qualitative agreement with those shown in Ref.~\cite{temp1}, with or without the 
contribution of TBF, which do not seem to have a major effect on the chemical potential. 

Next, we show the single-particle potential in SNM, $U(k,\rho,T)$, see Fig.~2. The effect of temperature is much more 
pronounced at low density (compare left and right panels) and low momenta. The general tendency is to turn slightly more
repulsive with increasing temperature, although this trend becomes clear only at the higher temperature. 
The increased repulsion is the result of a combination of effects. Temperature ``smears out" the step function
distribution, Eq.~(1), so that the interaction probability increases (decreases) at low (high) momenta due to 
the smaller (larger) occupation probability as compared to the $T=0$ case. Although Pauli blocking is 
generally reduced by the onset of temperature (which would suggest increased attraction among the particles), 
the integral in Eq.~(5) receives contributions from $G$-matrix elements at higher momenta (as compared                  
to the zero temperature case), and such contributions tend to be repulsive. 
 In the end, we observe a net effect that is repulsive, except at the lowest temperatures. 
Consistently with Fig.~2, the temperature dependence of the effective mass is generally small, see Fig.~3, 
and more noticeable at low density. Thus, in the range of densities and temperatures considered here, we expect only
minor effects on the in-medium cross sections, whose behavior is essentially dominated by the effective
mass. A slightly repulsive temperature effect on the DBHF effective mass was found in Ref.~\cite{HM87} as well.

The single-particle energy, not including the rest mass, is shown in Fig.~4 at fixed density and for different 
temperatures. Fig.~4 confirms that the single-particle interaction is noticeably impacted only at the lowest
momenta.                                                    
In Ref.~\cite{HM87} the equivalent Schroedinger optical
potentail was constructed from the Dirac self-energies and found to be remarkably insensitive to temperature.

\begin{figure}[!t] 
\centering          
\includegraphics[totalheight=2.7in]{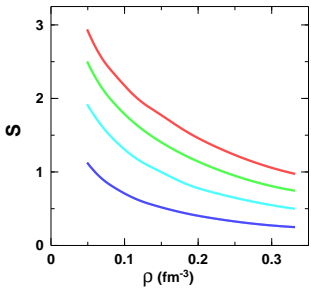}    
\vspace*{-1.5cm}
\caption{(color online)                                        
The entropy/particle in SNM as a function of density and  for increasing temperatures of $T=5$, $10$, $15$, and $20$ MeV. 
} 
\label{five}
\end{figure}

\begin{figure}[!t] 
\centering          
\includegraphics[totalheight=2.7in]{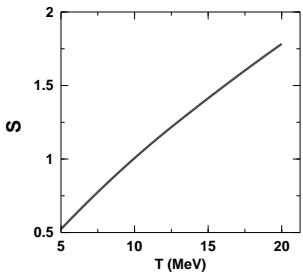}    
\vspace*{-1.5cm}
\caption{                                                      
The entropy/particle in SNM as a function of temperature at a density corresponding to a Fermi momentum of 1.3 $fm^{-1}$. 
} 
\label{six}
\end{figure}
We conclude this set of results by showing a macroscopic quantity, namely the entropy, which is a measure 
of thermal disorder. Entropy production in multifragmentation events in heavy-ion collisions is a crucial 
quantity in the determimation of the mass fragment distribution.                
The entropy per particle is shown in Fig.~5 as a function of density and for various temperatures, 
and in Fig.~6 it is displayed as a function of
temperature at a fixed density (close to saturation density).                      
The entropy increases with temperature, as physically reasonable, and decreases substantially with density.         
At low T, it is expected to approach a linear dependence due to the fact that,          
for a Fermi liquid, the relation between S and T should be approximately   
\begin{equation}
S \approx \frac{\pi ^2}{3\rho} N(T=0)T, 
\end{equation}
in terms of the density of states at the Fermi surface.                                                      

\begin{figure}[!t] 
\centering          
\includegraphics[totalheight=2.7in]{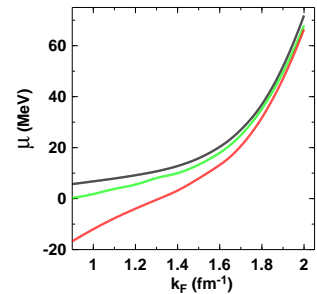}  
\vspace*{-1.2cm}
\caption{(color online)                                        
The chemical potential in neutron matter as a function of the Fermi momentum at $T=0$, $10$, and $20$ MeV. 
The chemical potential decreases with temperature. 
} 
\label{seven}
\end{figure}
\begin{figure}[!t] 
\centering          
\includegraphics[totalheight=2.7in]{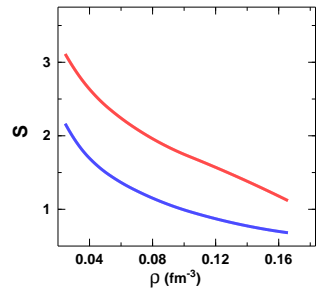}    
\vspace*{-1.5cm}
\caption{(color online)                                        
The entropy/particle in NM as a function of density at $T=10$ (lower curve) and $20$ MeV (upper curve). 
} 
\label{eight}
\end{figure}
\begin{figure}[!t] 
\centering          
\includegraphics[totalheight=2.7in]{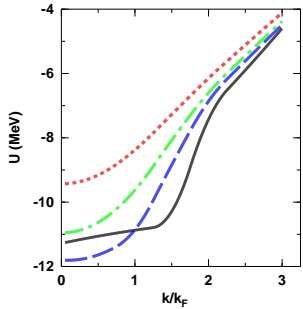}  
\vspace*{-1.2cm}
\caption{(color online)                                        
The single-particle potential in neutron matter as a function of the momentum at           
different temperatures: $T=0$ (solid black); $T=10$ (dashed blue); $T=20$ (dash-dotted green); $T=30$ (dotted red).
The neutron matter Fermi momentum is equal to 0.9 $fm^{-1}$.                  
} 
\label{nine}
\end{figure}
Although weak model dependence is not a general feature of thermodynamic quantities, 
the authors of Ref.~\cite{Rios06} demonstrate that different approximations to the entropy, including the 
quasi-particle approximation in the temperature dependent BHF scheme, differ from each other by 10 to 
20\% at most. This may be due to cancelations in the difference between the single-particle energy and the 
chemical potential \cite{Rios06}. 

\subsection{Neutron Matter} 
We have done a similar study of hot neutron matter (NM) as well. 
The chemical potential in NM is shown in Fig.~7 as a function of the Fermi momentum. (Of course, the same
Fermi momentum corresponds to half the density as compared to SNM.) The {\it microscopic} chemical potential,
as obtained from the normalization condition Eq.~(4), was shown to be in good agreement with the {\it 
macroscopic} one, obtained from the bulk properties through the derivative of the free energy density
\cite{Rios09}.                                                      

The entropy in NM is displayed in Fig.~8. As for the case of SNM, discrepancies between predictions from
different models have been found to be small, especially those arising from the use of different NN potential models
\cite{Rios09}. In fact, our predictions (based on the Bonn B potential) are close to those shown in
Ref.~\cite{Rios09} 
either with CDBONN \cite{CD} or Argonne V18 \cite{V18}. Concerning different many-body approaches, there are indications that contributions to the entropy from dynamical 
correlations (which would fragment the quasiparticle peak) are small. Hence, different 
predictions tend to approach the ``dynamical quasiparticle" result \cite{Rios09}.

The temperature dependence of the single-particle interaction in hot NM is comparable to,   
or smaller than, 
the one encountered in the SNM case. We show a representative case in Fig.~9. As discussed in Section {\bf IIIA}, we notice,
from Fig.~9 and from the left panel in Fig.~2, that the potential tends to become
deeper at first and then more shallow as the temperature increases. Also, temperature appears to
``wash out" some of the structure in the potential.

\section{Summary and conclusions}                                                                  

We used the Dirac-Brueckner-Hartree-Fock method extended to finite temperatures to 
predict single-particle properties in hot SNM and NM.                           
For temperatures up to a few tens of MeV the effect of temperature                                  
is small except at low densities and  momenta.                                                          
Due to suppression of Pauli blocking, which
is most important at low momentum, some temperature-induced modification of (low-energy) in-medium cross sections
could be expected. This is an interesting point to be explored further. 
We also discussed the entropy/particle. When compared with other predictions in the literature, our results
confirm the weak model dependence of this quantity. 

As usual, we adopt the microscopic approach for our nuclear matter calculations. 
Concerning our many-body method, we find              
DBHF to be a good starting point to look beyond the ground state of nuclear matter, which  it describes
successfully. The main strength of this method is its inherent ability to effectively incorporate 
crucial TBF contributions \cite{SL09}               
yet avoiding the possibility of inconsistency between the parameters of the two- and three-body systems. 

We have focussed on the one-body properties of the system, which are of great relevance for the study 
of energetic heavy-ion collision dynamics. 
A continuation of this study, including 
a close look at all thermodynamic quantities, will be the topic of a forthcoming work.                    

\section*{Acknowledgments}
Support from the U.S. Department of Energy under Grant No. DE-FG02-03ER41270 is 
acknowledged.


\newpage
\begin{references}             
\bibitem{Daniel05} P. Danielewicwz, nucl-th/0512009. 
\bibitem{MSU} B.-A. Li, P. Danielewicz, and W. Lynch, Phys. Rev. C {\bf 71}, 054603 (2005).
\bibitem{Zuo04} W. Zuo, Z.H. Li, A. Li, and G.C. Lu, arXiv:nucl-th/0412100, and references therein. 
\bibitem{SMN89} L. Satpathy, M. Mishra, and R. Nayak, Phys. Rev. C {\bf 39}, 162 (1989). 
\bibitem{JMZ83} H.R. Jaqaman, A.Z. Mekjian, and L. Zamick, Phys. Rev. C {\bf 27}, 2782 (1983); R.K. Su,
S.D. Yang, and T.T.S. Kuo, Phys. Rev. C {\bf 35}, 1539 (1987). 
\bibitem{Baldo95} M. Baldo, G. Giansiracusa, U. Lombardo, I. Bombaci, and L.S. Ferreira, Nucl. Phys. {\bf A583},
589c (1995). 
\bibitem{JMZ84} H.R. Jaqaman, A.Z. Mekjian, and L. Zamick, Phys. Rev. C {\bf 29}, 2067 (1984). 
\bibitem{HM87} B. ter Haar and R. Malfliet, Phys. Rep. {\bf 149}, 207 (1987).          
\bibitem{HWW98} H. Huber, F. Weber, and M.K. Weigel, Phys. Rev. C {\bf 57}, 3484 (1998). 
\bibitem{pol23}  A. Rios, A. Polls, and I. Vida{\~n}a, Phys. Rev. C {\bf 71}, 055802 (2005).
\bibitem{temp2}  I. Bombaci, A. Polls, A. Ramos, A. Rios, and I. Vida{\~n}a, nucl-th/0506016.                       
\bibitem{V18} R.B. Wiringa, V.G.J. Stocks, and R. Schiavilla, Phys. Rev. C {\bf 51}, 38 (1995). 
\bibitem{Lopez06} D. Lopez-Val, A. Rios, A. Polls, and I. Vida{\~n}a, Phys. Rev C {\bf 74}, 068801 (2006). 
\bibitem{temp1} M. Baldo and L.S. Ferreira, Phys. Rev. C {\bf 59}, 682 (1999).
\bibitem{FM03} T. Frick and H. M{\"u}ther, Phys. Rev. C {\bf 68}, 034310 (2003). 
\bibitem{Rios06} A. Rios, A. Polls, A. Ramos, and H. M{\"u}ther, Phys. Rev. C {\bf 74}, 054317 (2006). 
\bibitem{Rios09}  A. Rios, A. Polls, and I. Vida{\~n}a, Phys. Rev. C {\bf 79}, 025802 (2009).
\bibitem{Moust08}  Ch. C. Moustakidis, Phys. Rev. C {\bf 78}, 054323 (2008).
\bibitem{Moust09}  Ch. C. Moustakidis and C.P. Panos, Phys. Rev. C {\bf 79}, 045806 (2009).
\bibitem{AS03} D. Alonso and F. Sammarruca, Phys. Rev. C {\bf 67}, 054301 (2003). 
\bibitem{SL08} F. Sammarruca and Pei Liu, arXiv:0806.1936 [nucl-th].
\bibitem{SL09} F. Sammarruca and Pei Liu, arXiv:0906.0320 [nucl-th].
\bibitem{SK07} F. Sammarruca and P. Krastev, Phys. Rev. C {\bf 75}, 034315 (2007). 
\bibitem{Mac89} R. Machleidt, Adv. Nucl. Phys. {\bf 19}, 189 (1989). 
\bibitem{Das+} C. Das and R.K. Tripathi, Phys. Rev. C {\bf 45}, 2217 (1992). 
\bibitem{CD} R. Machleidt, Phys. Rev. C {\bf 63}, 024001 (2001). 
\end{references}
\end{document}